\documentstyle{prhep97}

\makeatletter
\let\chapter\hid@chapter
\makeatother
\begin{document}


\authorrunning{S.\,Bellucci}
\titlerunning{{\talknumber}: Antigravity phenomenology}
 

\def\talknumber{1260} 

\title{{\talknumber}: Phenomenology of Antigravity in $N=2,8$ Supergravity}
\author{Stefano\, Bellucci
(bellucci@lnf.infn.it)}
\institute{INFN-Laboratori Nazionali di Frascati, Frascati, Italy}

\maketitle

\begin{abstract}
$N=2,8$ supergravity predicts antigravity (gravivector
and graviscalar) fields in the graviton supermultiplet. Data on the
binary pulsar PSR 1913+16, tests of the equivalence principle and
searches for a fifth force yield an upper bound of order 1 meter
(respectively, 100 meters) on the range of the gravivector
(respectively, graviscalar) interaction. Hence these fields are not
important in non-relativistic astrophysics (for the weak-field limit of
$N=2,8$ supergravity) but can play a role near black holes and
for primordial structures in the early universe of a size comparable to
their Compton wavelengths.
\end{abstract}
%
The quest for a unified description of elementary particle and gravity
theories led to local supersymmetry \cite{pvn}. The large symmetry content
of supergravity yields, in spite of its lack of renormalizability,
powerful constraints on physical observables, e.g. anomalous magnetic
moments \cite{g2}.

It has been shown that a clear case for antigravity theories arises,
when considering $N>1$ supergravity theories \cite{Scherk,ScherkProc}.
Combining laboratory data together with
geophysical and astronomical observations has
provided us restrictions on the
antigravity features of some extended supergravity 
theories \cite{belfar,bf2}.
This can have important consequences for high precision experiments
measuring the difference in the gravitational acceleration of
the proton and the antiproton \cite{PS-200}.
A review of earlier ideas about antigravity
is found in \cite{GoldmanNieto}.
 
The $N=2,8$ supergravity multiplets contain, in addition to
the graviton, a vector field $A_{\mu}^l$ \cite{Zachos},
\cite{SS,CremmerSS}.
This field, which we refer to as the gravivector, carries antigravity,
because it couples to quarks and leptons with a positive sign
and to antiquark and antileptons with a negative one. The coupling is
proportional to the mass of the matter fields and vanishes
for self-conjugated particles.
The other antigravity field is the scalar $\sigma$ entering the $N=8$
supergravity multiplet \cite{Scherk,ScherkProc}. We refer to it as
the graviscalar.

We are bound, in force of the result of the E\"otv\"os experiment,
to take a nonvanishing mass for the field
$A_{\mu}^l$ \cite{Scherk,ScherkProc}.
\begin{equation}
m_l=\frac{1}{R_l}=\sqrt{4\pi G_N}\, m_{\phi}\langle \phi \rangle \; ,
m_{\phi}=\langle \phi \rangle \, ,      
\end{equation}
where the Higgs mechanism has been invoked.

The presence of the gravivector in the theory introduces a violation of
the equivalence principle on a range of distances of order
the Compton wavelength $R_l$.
At present, the equivalence principle is verified with a precision
$|\delta \gamma$/$\gamma | < 3\times 10^{-12}$ \cite{EotWash}.\footnote{The
equivalence principle is advocated in a recent proposal of a consistent
quantum gravity \cite{bs}.}
This number was used in \cite{bf2}, in order
to constrain the {\em v.e.v.} of the scalar field $\phi$, and
therefore its mass
$m_{\phi}>15$ (31) GeV, N=2 (8).   
The above constraint on the field that gives to the gravivector its mass
corresponds to an upper bound of order 1 m for $R_l$
\cite{bf2}.

It is worth to remind the reader that there are interesting 
connections between antigravity in $N=2,8$ supergravities and {\em 
CP} violation experiments, via the consideration of the 
$K^0$--$\overline{K}^0$ system in the terrestrial gravitational field 
\cite{Scherk}.\footnote{For a low-energy theorem in gravity coupled to
scalar matter, see e.g. \cite{abs}.}
However, the present experiments on {\em CP} violations 
yield bounds on the range of the gravivector field which are less 
stringent than those obtained from the tests of the equivalence 
principle \cite{belfar}.

The null results of the search for possible
deviations from Newton's law reported
in \cite{Spero} forbid values in the following ranges
\cite{bf2}:
\begin{equation}
82 \: \mbox{GeV} <m_{\phi}<
376 \: \mbox{GeV} \:\:\:\:\:\:\:\: (N=2)\; , 
\end{equation}
\begin{equation}
46 \: \mbox{GeV }<m_{\phi}<
461 \; \mbox{GeV} \:\:\:\:\:\:\:\: (N=8) \; .
\end{equation}

A high precision test of the equivalence principle in the 
field of the Earth is currently under planning in Moscow \cite{Kalebin}.
The precision expected to be achieved in this experiment is 
$|\delta \gamma$/$\gamma | < 3\times 10^{-15}$.
In the case that the new experiment verifies the equivalence principle 
with the expected accuracy, the limits on $m_{\phi}$ would be pushed 
to
$m_{\phi}>0.5$ (1) TeV, $N=2$ (8).
\footnote{
A caveat concerning our results on the gravivector is that
the presence of a $U(1)$ symmetry for the $D=4$ extended
supergravity theory obtained
by dimensional reduction from a higher dimension implies a mass
for this field of order the Planck mass \cite{SS}.
In this particular instance it is unlikely that experimental
limits on the gravivector have any physical application
(aside perhaps from applications to inflationary models, if it were
possible to use a vector field instead of a scalar). We plan to come back
to this issue in a further study.}

If supersymmetry is unbroken, the violation of the equivalence
principle due to the graviscalar of $N=8$ supergravity takes the form of
a universal spatial dependence in the effective (gravitational) Newton's
coupling, $G_N=G_N(r)$ \cite{ScherkProc}.
However there are corrections due to the breaking of supersymmetry -
which we hope to account for in a forthcoming publication -
and those depend on the composition of the material. If we neglect
them for the time being, then there is no effect of $\sigma$
in  E\"otv\"os-like experiments, where the acceleration difference between
two bodies of different composition is measured. In this case it is
still possible to constrain the effective range of the
$\sigma$-mediated interaction, analyzing data from the binary pulsar
PSR 1913+16 \cite{bi}, constraints from the observations searching for
a fifth force \cite{fifth}, and some of the experiments aimed at testing
Newton's inverse square law \cite{Spero}. In this way we
got the following bounds on the Compton
wavelength of the graviscalar:
$R_{\sigma}<0.15$ cm, 70 m$<R_{\sigma}<100$ m
\cite{bf2}.

There have been many papers on the effects of non-Newtonian gravity in
astrophysics, in particular those due to a fifth force like the one obtainable
from $N=2,8$ supergravity in the weak field limit (see references in
\cite{GoldmanNieto}).\footnote{The effect of the cosmological constant
is shadowed by Newtonian-gravity effects \cite{odin,ads}. Large-scale
topology-changing configurations can suppress its effect also for
1-loop scalar-$QED$, see e.g. \cite{bd}.}
However, the upper bounds of order 1 m (100 m) on the Compton wavelength $R_l$
($R_{\sigma}$) of 
the gravivector (graviscalar) field found in \cite{bf2} imply that 
antigravity effects induced by the extended supergravity theories
do not play any role in nonrelativistic astrophysics, since 
the length scales involved in stellar,\footnote{The conclusion that the
stellar 
structure is unaffected by antigravity might change if the non-Newtonian 
force alters the equation of state of the matter composing the star 
\cite{stellar}.} galactic and supergalactic 
structures dominated by gravity are much larger than $R_l$ and ($R_{\sigma}$). 

This supergravity-induced
antigravity could affect, in principle, processes that take place in the 
strong gravity regime, where smaller distance scales are involved. 
Examples of these situations are processes occurring near black hole 
horizons or in the early universe, when the size of the universe is 
smaller than, or of the order of, $R_l$ and ($R_{\sigma}$).
The relevance of the supergravity-induced antigravity 
in such situations will be studied in future publications.

Our final remark concerns a point that apparently went unnoticed in 
the literature on supergravity: the detection of gravitational waves 
expected in a not too far future will shed light on the 
correctness of supergravity theories. In fact, after the dimensional 
reduction is performed, the action of the theory contains scalar and 
vector fields as well as the usual metric tensor associated to spin~2 
gravitons \cite{ScherkProc}. These fields are responsible for the presence 
of polarization modes in gravitational waves, whose effect differs from 
that of the spin~2 modes familiar from general 
relativity. Therefore, extended supergravities and general 
relativity occupy different classes in the $E(2)$ classification
of gravity theories \cite{Eardleyetal}. The extra 
polarization states are detectable, in principle, in a gravitational 
wave experiment employing a suitable array of detectors \cite{Eardleyetal}. 
However, it must be noted that a detailed study of gravitational wave 
generation  taking into account the antigravity phenomenon is not available
at present. Such a work would undoubtedly have to face the remarkable 
difficulties well known from the studies of gravitational wave generation
in the context of general relativity.

We are grateful to G.A. Lobov for drawing our attention to the ITEP
experiment. We acknowledge useful comments by C. Kounnas and G. Veneziano.
                              
\end{document}